# Controlling shot noise in double-barrier magnetic tunnel junctions

J.P. Cascales,[1] D. Herranz,[1] F. G. Aliev*,[1] T. Szczepański,[2] V. K. Dugaev,[2,3] J. Barnaś,[4,5] A. Duluard,[6] M. Hehn,[6] and C. Tiusan[6]

[1]*Depto. Fisica Materia Condensada, C03, Universidad Autonoma de Madrid, 28049, Madrid, Spain*
[2]*Department of Physics, Rzeszów University of Technology, 35-959 Rzeszów, Poland*
[3]*Institut für Physik, Martin-Luther-Universität Halle-Wittenberg, 06120 Halle, Germany*
[4]*Faculty of Physics, Adam Mickiewicz University, 61-614 Poznań, Poland*
[5]*Institute of Molecular Physics, Polish Academy of Sciences, 60-179 Poznań, Poland*
[6]*Institute Jean Lamour, Nancy Universitè, 54506 Vandoeuvre-les-Nancy Cedex, France*
(Dated: September 20, 2012)

## Abstract

We demonstrate that shot noise in Fe/MgO/Fe/MgO/Fe double-barrier magnetic tunnel junctions is determined by the relative magnetic configuration of the junction and also by the asymmetry of the barriers. The proposed theoretical model, based on sequential tunneling through the system and including spin relaxation, successfully accounts for the experimental observations for bias voltages below 0.5V, where the influence of quantum well states is negligible. A weak enhancement of conductance and shot noise, observed at some voltages (especially above 0.5V), indicates the formation of quantum well states in the middle magnetic layer. The observed results open up new perspectives for a reliable magnetic control of the most fundamental noise in spintronic structures.

As solid-state electronic devices shrink in size, further advances essentially depend on the understanding and control of spontaneous off-equilibrium fluctuations in charge and/or spin currents. Being a consequence of the discrete nature of charge carriers, shot noise (SN) is the only contribution to the noise which survives at low temperatures. Moreover, SN is an excellent tool to investigate the correlations and coherency at the nanoscale, well beyond the capabilities of electron transport[1–9]. In the absence of correlations, SN is Poissonian (full shot noise) and its noise power is given by $S_{full} = 2eI$, where $I$ is the average current and $e$ the electron charge. The Fano factor, $F = S_{exp}/S_{full}$, represents the experimental SN normalized by the full SN value. It is generally suppressed ($F < 1$) by electron correlations[1] (quantum and/or Coulomb), but it can also be enhanced ($F > 1$), e.g. due to tunneling via localized states[10].

After the observation of spin dependent transport in Fe/MgO/Fe magnetic tunnel junctions (MTJs)[11,12], MgO-based junctions became important elements of spintronic devices. Moreover, the recent implementation of MgO for an effective spin injection[13,14] revealed a new road for reducing the spin relaxation due to conductivity mismatch[15,16]. The efforts aimed at understanding spin coherency and SN, limited up to now to MTJs, revealed suppressed SN with $Al_2O_3$ barriers ($0.7 < F < 1$) due to sequential tunneling[17] and also in serial MTJ arrays[18]. As for MTJs with MgO barriers, full SN ($F = 1$) independent of the magnetic state was observed in epitaxial Fe/MgO/Fe[19]. Then, the noise was examined for ultra-thin (less than 1 nm) MgO barriers, where $F \simeq 0.92$ was observed in the parallel state[20,21].

Double-barrier magnetic tunnel junctions (DMTJs), with either nanoparticles[22,23] or a continuous magnetic layer as the central electrode[24], have some advantages in comparison with MTJs. First, they show an enhanced tunnel magnetoresistance (TMR)[24,25], which additionally reveals oscillations induced by quantum well states (QWSs)[23,26]. Second, spin accumulation in the central layer is expected to substantially enhance spin torque[27,28]. The investigation of the statistics of spin tunneling events in hybrid spintronic devices is of great potential interest also beyond the spintronics community. From a general point of view, an experiment that measures SN in a structure with three magnetic layers in DMTJs is similar to an experiment on photon statistics in a device with one polarizer and two analyzers[29]. From a practical point of view, as we demonstrate below, DMTJs are unique devices which allow to engineer and control the most fundamental noise mechanism by simply switching the device between its different magnetic states.

In this Letter we report on the investigation of shot noise in seminal, epitaxial Fe/MgO/Fe/MgO/Fe DMTJs. We show that SN can be controlled by the magnetic state of DMTJs and also by the asymmetry of the two MgO barriers. The measurements at biases where the influence of QWS is small are in good agreement with the model, which takes into account spin relaxation in the central electrode. Our findings reveal new perspectives for the magnetic control of SN and also present a novel method to quantify the electron spin relaxation in spintronic devices.

Three different types DMTJs were grown at room temperature by Molecular Beam Epitaxy (MBE) on MgO (100) substrates under ultra high vacuum (UHV) conditions. The first type of junctions (DMTJ1), with strongly asymmetric barriers, have the structure: MgO//MgO(15)/Fe$_1$(45)/MgO(11ML) /Fe$_2$(10)/MgO(3ML)/Fe$_3$(20)/Au(20) (numbers in parenthesis are thicknesses in nm, while ML stands for monolayer). Type 2 (DMTJ2) and type 3 (DMTJ3) junctions are composed of the layers MgO//MgO(10)/Cr(42)/Co(10)/Fe$_1$(5)/MgO(10 and 8ML)/Fe$_2$(5)/MgO(9ML)/Fe$_3$(10)/Co(30)/Au(10) and are more symmetric. The barrier asymmetry of DMTJ2 samples is opposite to that of DMTJ3. We label the topmost barrier as number 1 and the bottom barrier as 2. The wafers for DMTJ2 and DMTJ3 were grown simultaneously. The 2ML difference in thickness of the bottom barrier is controlled by a shutter. Monolayer-level control has been achieved by monitoring in-situ the intensity oscillations in the reflection high-energy electron diffraction (RHEED) pattern along the [100] direction during the layer-by-layer growth of the MgO barriers. Square shaped MTJs and DMTJs with lateral sizes from 10 to 30 $\mu m$ have been patterned by an optical lithography/ion etching process, controlled by Auger spectroscopy[24]. The setup for conductance and SN measurements was described earlier[17,19]. Although the SN and electron transport measurements were done at 0.3K and 4K, at the highest biases the real sample temperature was below 10K, estimated by comparing the $I-V$ curves measured at 0.3K, 4K and 10K.

Figure 1 compares the zero bias TMR in the three types of DMTJs. In agreement with previous findings[24,25], the antiferromagnetic coupling between two ferromagnetic layers across the thinnest MgO barrier in DMTJ1 results in the presence of two different AP1 ($\uparrow \downarrow\uparrow$) and AP1$^r$ ($\downarrow \uparrow\downarrow$) states for which the central layer is aligned opposite to the neighboring ones. We attribute the difference in resistance between the AP1 and AP1$^r$ states to the possible influence of domain walls formed in the synthetic antiferromagnet in the latter state, suggested by the large 1/f noise observed. The TMR dip in-between corresponds to the AP2 ($\downarrow \downarrow\uparrow$) state. Both barriers in DMTJ2 and DMTJ3 junctions differ only slightly, thus providing lower barrier asymmetries. As expected from the dependence of the TMR effect on the barrier thickness in Fe/MgO MTJs, DMTJ2s show a larger resistance jump between the AP1 ($\uparrow \downarrow \uparrow$) and AP2 ($\downarrow \downarrow \uparrow$) states than between the AP2 ($\downarrow \downarrow \uparrow$) and P ($\downarrow \downarrow \downarrow$) states. DMTJ3 samples show the opposite behavior.

Figures 2 and 3 present our main experimental findings: the suppression of shot noise below the classic ($F = 1$) value in weakly asymmetric DMTJs. The Fano factor was obtained by normalizing the experimentally measured SN at fixed T by the full shot noise at the same T. This approximation is justified above 100mV, where $eV/k_BT > 100$. Depending on the magnetic state and bias, the Fano factor varies in the range of $F \in (0.5-0.9)$. We also observed that SN is only weakly suppressed ($F \sim 0.9$, see below) and is nearly independent of the magnetic state in the DMTJ1 junctions. In agreement with previous reports[19], SN has a nearly Poissonian character ($F \simeq 1$) in epitaxial MTJs with a single (2.5nm MgO) barrier (not shown). Solid curves in Fig 2 show the estimated full SN from the $I-V$ curves at T=0.3K:

$S_V = 2eI/G_d^2$, where $G_d$ is the differential conductance.

The differential conductance (after substracting the parabolic background $G_0$) and the Fano factor for DMTJ2 are compared in Fig.3. The well defined oscillations of the differential conductance indicate the presence of resonant transmission through QWSs formed in the central Fe$_2$ layer. The observed oscillation period is around 300 mV and is in rough agreement with the predictions for DMTJs with a 5nm central layer[23]. QWSs are more pronounced for positive biases (when electrons tunnel from the bottom to the upper electrodes) for DMTJs1,2, and, as could be expected from the barrier asymmetry, are more pronounced for negative bias for DMTJ3. Indeed, the voltage distribution across an asymetric DMTJ shows that QWSs affect the conductance mainly when electrons tunnel from the contact's Fermi level to the central layer through the thicker barrier[30].

Depending on the degree of coherency involved in the transmission through QWSs, SN is expected to show a shallow dip in the Fano factor due to coherent resonant transmission followed by a resonant enhancement in the negative differential conductance regime due to Coulomb interactions[1,10,31]. These anomalies in Fano are more pronounced for DMTJ2 and for positive bias, where QWSs in the conductance are more clearly observed (Fig.3). We also remark that the observed periodic anomalies in the conductance and Fano factor cannot be attributed to Coulomb blockade effects, where the Fano factor has periodic minima decreasing in amplitude with the applied bias[32].

In order to understand the variation of the Fano factor with the barrier asymmetry and magnetic state of the system, we calculate the shot noise using a model of sequential tunneling without taking into consideration the influence of resonant tunneling. In the absence of spin relaxation, the shot noise power $S$ can be calculated as $S = S_\uparrow + S_\downarrow$, where $S_\uparrow$ and $S_\downarrow$ are the contributions from the two separate spin-up and spin-down channels. The noise power $S_\sigma$ is then exactly like in the case of spinless particles,[33]

$$S_\sigma = \left(R_{1\sigma}^2 S_{1\sigma} + R_{2\sigma}^2 S_{2\sigma}\right)/R_\sigma^2, \quad (1)$$

where $R_\sigma = R_{1\sigma} + R_{2\sigma}$ with $R_{i\sigma}$ being the spin dependent resistance of the $i$-th barrier ($i = 1, 2$), while $S_{i\sigma} = 2eV/R_\sigma$. The above relation holds for small transmission through the barriers. Therefore, the Fano factor, $F = S/2eI$, can be calculated as

$$F = \frac{(R_{1\uparrow}^2 + R_{2\uparrow}^2)R_\downarrow^3 + (R_{1\downarrow}^2 + R_{2\downarrow}^2)R_\uparrow^3}{R_\uparrow^2 R_\downarrow^2 (R_\uparrow + R_\downarrow)}. \quad (2)$$

We introduce the following parameters: $\alpha = R_{2\uparrow}^0/R_{1\uparrow}^0$, $\beta_1 = R_{1\downarrow}^0/R_{1\uparrow}^0$, and $\beta_2 = R_{2\downarrow}^0/R_{2\uparrow}^0$, where $R_{i\uparrow(\downarrow)}^0$ is the barrier resistance for spin majority (minority) electrons in the state with parallel magnetizations on both sides of the $i$-th barrier. Thus, in the P configuration $R_{i\sigma} = R_{i\sigma}^0$, while in the AP1 configuration ($\uparrow\downarrow\uparrow$) one finds $R_{i\sigma} =$ $\sqrt{R_{i\uparrow}^0 R_{i\downarrow}^0}$, and for the AP2 configuration ($\downarrow\downarrow\uparrow$) one can write $R_{2\sigma} = R_{2\sigma}^0$ and $R_{1\sigma} = \sqrt{R_{1\uparrow}^0 R_{1\downarrow}^0}$. In the symmetric case, $\alpha = 1$ and $\beta_1 = \beta_2 = \beta$, the above results lead to $F = 1/2$ for the P and AP1 configurations, and

$$F = \frac{1+\beta}{(1+\sqrt{\beta})^2} \quad (3)$$

in the AP2 configuration.

The above simplified approach neglecting spin relaxation in the central layer, however, cannot account for the main experimental observations. Therefore, we now take into account the spin relaxation and write the relevant equation for spin density fluctuations $\delta S_z$,

$$\Delta J_z^{(2)} - \Delta J_z^{(1)} = -\frac{\delta S_z}{\tau_s}, \quad (4)$$

where $J_z^{(i)}$ is the $z$-component of spin current in the $i$-th barrier and $\tau_s$ is the spin relaxation time in the central electrode. As shown below, the experimental data can be accounted for rather well with a relatively short spin relaxation time. In the limit of strong spin relaxation one can completely neglect the spin fluctuations and take into account only the charge fluctuations. Instead of Eq. (2) one then finds

$$F = \frac{R_{2\uparrow} R_{2\downarrow} (R_{1\uparrow} + R_{1\downarrow})^2 + R_{1\uparrow} R_{1\downarrow} (R_{2\uparrow} + R_{2\downarrow})^2}{[R_{1\uparrow} R_{1\downarrow} (R_{2\uparrow} + R_{2\downarrow}) + R_{2\uparrow} R_{2\downarrow} (R_{1\uparrow} + R_{1\downarrow})]^2}. \quad (5)$$

From this formula one can calculate the Fano factors in the P, AP1 and AP2 configurations, similarly as in the case without spin relaxation. In a general case, the spin fluctuations have been taken into account *via* Eq.(4). The corresponding formulas, however, are cumbersome and will not be presented here.

To compare the theoretical results with the experimental data we used average Fano values for the biases between 0.2 and 0.5 V in order to avoid the possible influence of defect states in the barrier below 200 mV[34], and to minimize the influence of QWSs observed mainly above 0.5V. Figures 4(a-c) show the calculated Fano factors as a function of the asymmetry parameter $\alpha$ for all three states, together with the experimental Fano values for DMTJ1-3. There is a good qualitative and quantitative agreement with the experimental results. We see that the combined TMR and SN provide an evaluation of the three independent parameters $\alpha$, $\beta_1$ and $\beta_2$.

The key element of the theory is the dependence of the SN on the spin density fluctuations. These fluctuations are described by the kinetic equation (4) and depend on the spin relaxation (described conveniently by the parameter $g = d/v_F \tau_s$, with $d$ being the thickness of the central layer, and $v_F$ the Fermi velocity). Figure 4d shows the estimated parameter $g$ and the barrier asymmetry $\alpha$ for our DMTJs. It is interesting to note that the best fits to the theory for two measured DMTJ3s (see Fig.4c) appear with relatively low $g$ (i.e. large $\tau_s$, estimated to be around $10^{-12}$s for $v_F = 10^4$m/s). On the other hand, SN in both measured DMTJ2s is best described with $g \sim 100$ (i.e. short $\tau_s$) as seen from Fig.4d. We relate shorter $\tau_s$ in DMTJ2s with an increased density of oblique defects as the epitaxial MgO is grown above the critical thickness for the plastic relaxation of MgO on Fe[24]. These defects could be "imprinted" on the central electrode, increasing its defectiveness and, in agreement with the Overhauser-Elliott-Yafet model[35], strongly reducing $\tau_s$.

We note that the model neglects other possible sources of the noise, like $1/f$ and thermal noise. Apart from this, the SN is calculated when neglecting spin coherent resonant tunneling. Moreover, our model does not include any deviation of the angle between magnetizations from 0 or $\pi$, which may influence the Fano factor[3]. We also omitted the influence of disorder and interfacial states, which may reduce the Fano factor[21]. All these factors can be responsible for the deviation of the theoretical curves from the experimental points in Fig.4(a-c). The strongest deviation in the case of DMTJ1 (Fig.4a) could be attributed to the presence of exchange coupling for thin (3mL) MgO barriers[24,25], with the possible formation of domain walls in the central Fe electrode.

In conclusion, we have demonstrated that the shot noise in DMTJs with low barrier asymmetry can be effectively reduced below the full shot noise value. Furthermore, SN is influenced by the relative magnetic configuration of a DMTJ and also depends on the bias. Moreover, our work presents a novel method to study the spin relaxation time in the central electrode of a DMTJ using SN measurements. The capability to reduce the most fundamental noise source in electronics could be useful both for vertical (e.g. spin current injection in semiconductors through double MgO barriers) or lateral (e.g. quantum dots) electronic structures.

The authors thank A. Gomez-Ibarlucea and C. C. Bellouard for their help with the experiment. This work was supported by the Spanish MICINN (MAT2009-10139, CSD2007-00010, FR2009-0010 grants), Communidad de Madrid (P2009/MAT-1726) and by the DFG in Germany and by the NCN in Poland as a research project for the years 2011-2014.

---

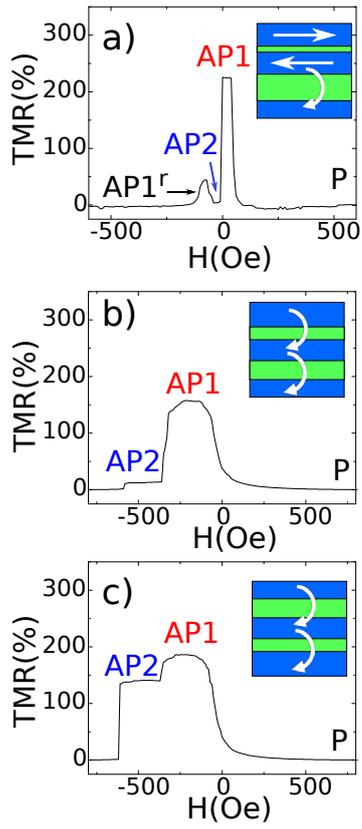

FIG. 1: Tunneling magnetoresistance in DMTJs at T=4K with different barrier asymmetries. (a) TMR in DMTJ1 with one very thin barrier, resulting in the coupling of the top and central magnetic layers. (b) TMR in DMTJ2 junctions with a thick barrier 2. The resistance difference between AP1 and AP2 is higher than between AP2 and P. (c) TMR in DMTJ3 with a thick barrier 1. This results in a bigger resistance jump from AP2 to P, than from AP1 to AP2.

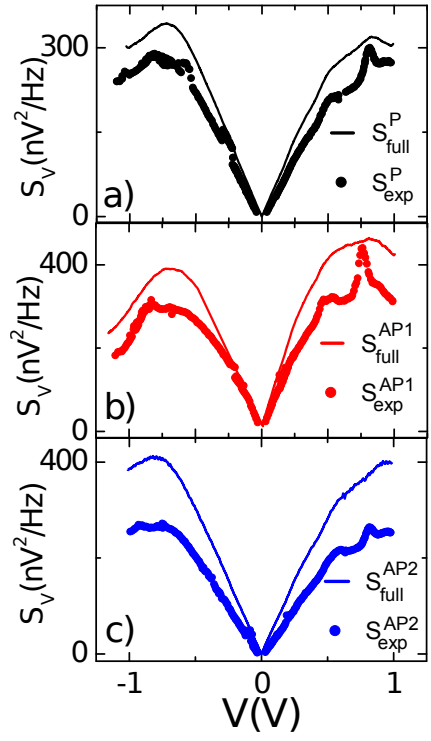

FIG. 2: Typical bias dependence of the shot noise for DMTJ2 measured in three different magnetic configurations at T=0.3K. The experimental data (points) is compared to full shot noise (lines).

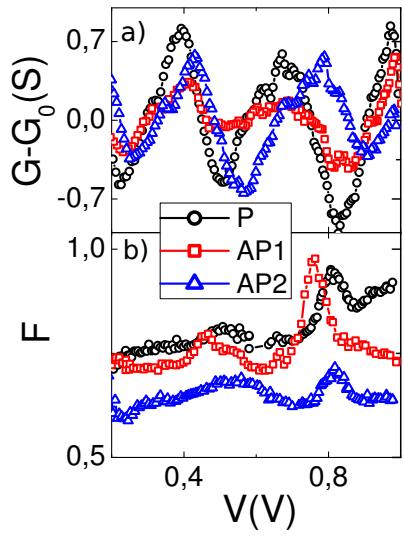

FIG. 3: (a) Bias dependence of the dynamic conductance (after the subtraction of a parabolic background) for three different magnetic states at $T = 4K$. (b) Bias dependence of the Fano factor measured in the corresponding three magnetic states of a DMTJ2 sample

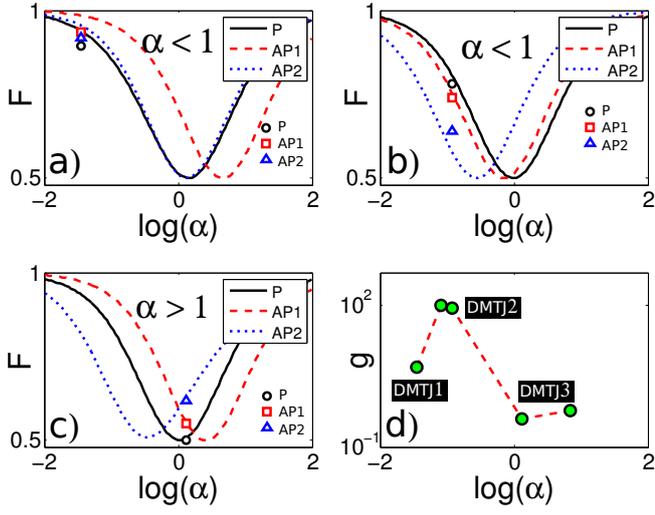

FIG. 4: Graphs (a-c) compare the theory (solid lines) with the experimental values (points) of the Fano factor measured for three different DMTJs. (a) DMTJ1 ($\alpha = 0.035$, $g = 4.9$, $\beta_1 = 42$, $\beta_2 = 2$) (b) DMTJ2 ($\alpha = 0.12$, $g = 90$, $\beta_1 = 23$, $\beta_2 = 48$); (c) DMTJ3 ($\alpha = 1.3$, $g = 0.3$, $\beta_1 = 75$, $\beta_2 = 11$). Part (d) presents the minimum values of the spin relaxation parameter $g$ which gives correct shot noise and TMR values as a function of the barrier asymmetry.